\documentclass[prl,aps,showpacs,twocolumn,preprintnumbers,amsmath,amssymb,superscriptaddress,longbibliography]{revtex4-2}
\usepackage[english]{babel}
\usepackage{amsmath,amssymb,bbm,mathrsfs,mathtools,multirow,diagbox,xcolor,%
  graphicx,tabularx,comment,amsfonts,dsfont,lettrine,units,comment}
\usepackage{graphicx}
\usepackage[colorlinks=True,linkcolor=red,citecolor=blue,urlcolor=blue]{hyperref}
\usepackage{bookmark}
\usepackage{xcolor}
\usepackage{chngcntr}
\usepackage{braket}
\usepackage{nicefrac}
\usepackage{bm}
\usepackage{bbm}
\usepackage{braket}
\usepackage{lineno}
\usepackage{array}
\newcolumntype{C}{>{$}c<{$}}
\usepackage{newtxtext} 
\usepackage{ulem}[normalem]   
\usepackage{orcidlink}

\usepackage{graphicx}
\usepackage{bm}

\usepackage{hyperref}
\hypersetup{
    colorlinks=true,
    linkcolor=blue,
    filecolor=blue,      
    urlcolor=blue,
    citecolor=blue,
    pdftitle={a-Kitaev},    
    pdfkeywords={amorphous}, 
}
 
\makeatletter
\renewcommand*\env@matrix[1][*\c@MaxMatrixCols c]{%
  \hskip -\arraycolsep
  \let\@ifnextchar\new@ifnextchar
  \array{#1}}
\makeatother

\begin{document}

\title{
Amorphous and polycrystalline routes towards a chiral spin liquid
}

\author{Adolfo G. Grushin\orcidlink{0000-0001-7678-7100}}
\affiliation{Univ. Grenoble Alpes, CNRS, Grenoble INP, Institut Néel, 38000 Grenoble, France}
\author{C\'{e}cile Repellin\orcidlink{0000-0003-2420-6815}} 
\affiliation{Univ. Grenoble Alpes, CNRS, Grenoble INP, LPMMC, 38000 Grenoble, France}

\date{\today}

\begin{abstract}
We show that a chiral spin liquid spontaneously emerges in partially amorphous, polycrystalline,  or ion-irradiated Kitaev materials.
In these systems, time-reversal symmetry is broken spontaneously due to a non-zero density of plaquettes with an odd number of edges, $n_{odd}$. This mechanism opens a sizable gap, at small $n_{odd}$ compatible with that of typical amorphous materials and polycrystals, and which can alternatively be induced by ion irradiation. We find that the gap is proportional to $n_{odd}$, saturating at $n_{odd}\sim 40\%$. Using exact diagonalization, we find that the chiral spin liquid is approximately as stable to Heisenberg interactions as Kitaev's honeycomb spin-liquid model. Our results open up a significant number of noncrystalline systems where chiral spin liquids can emerge without external magnetic fields.
\end{abstract}

\maketitle

The search for topological phases and materials has focused on crystals due to the convenience of translational symmetry to calculate topological invariants. Hence, amorphous solids represent the largest subset of materials that remain unclassified in terms of their topological properties~\cite{Zallen,Grushin2020,Corbae2023}. This observation opens a large material class~\cite{Corbae:2019tg} to search for topological phenomena. Their phenomenology can parallel that of crystals, as all strong topological phases can exist in amorphous lattices~\cite{Agarwala:2017jv,Mitchell2018,Poyhonen2017,Bourne:2018jr,manna2022noncrystalline}, but also extend it, for example due to the presence of average symmetries~\cite{Spring2021,Marsal2022}.
In that sense,  amorphous topological phases add to the zoo of topological phases that exist because of disorder, rather than in spite of it~\cite{Li2009,Jiang2009,Groth09,Agarwala2020,Kim2023}. 

Amorphous materials could bring a new perspective on other fundamental open problems, such as the challenge of finding quantum spin-liquid materials. Quantum spin liquids are entangled phases of matter characterized by the absence of symmetry breaking at zero temperature. The search for new candidate materials displaying clear spin-liquid signatures remains a central goal, despite encouraging experimental evidence~\cite{lee_review_2008, Savary2016,knolle_moessner_review_2019}. As in the search of new topological phases and materials, amorphous solids represent a new pool of available materials and new physical properties.
\begin{figure}[ht!]
    \centering
    \includegraphics[width=\columnwidth]{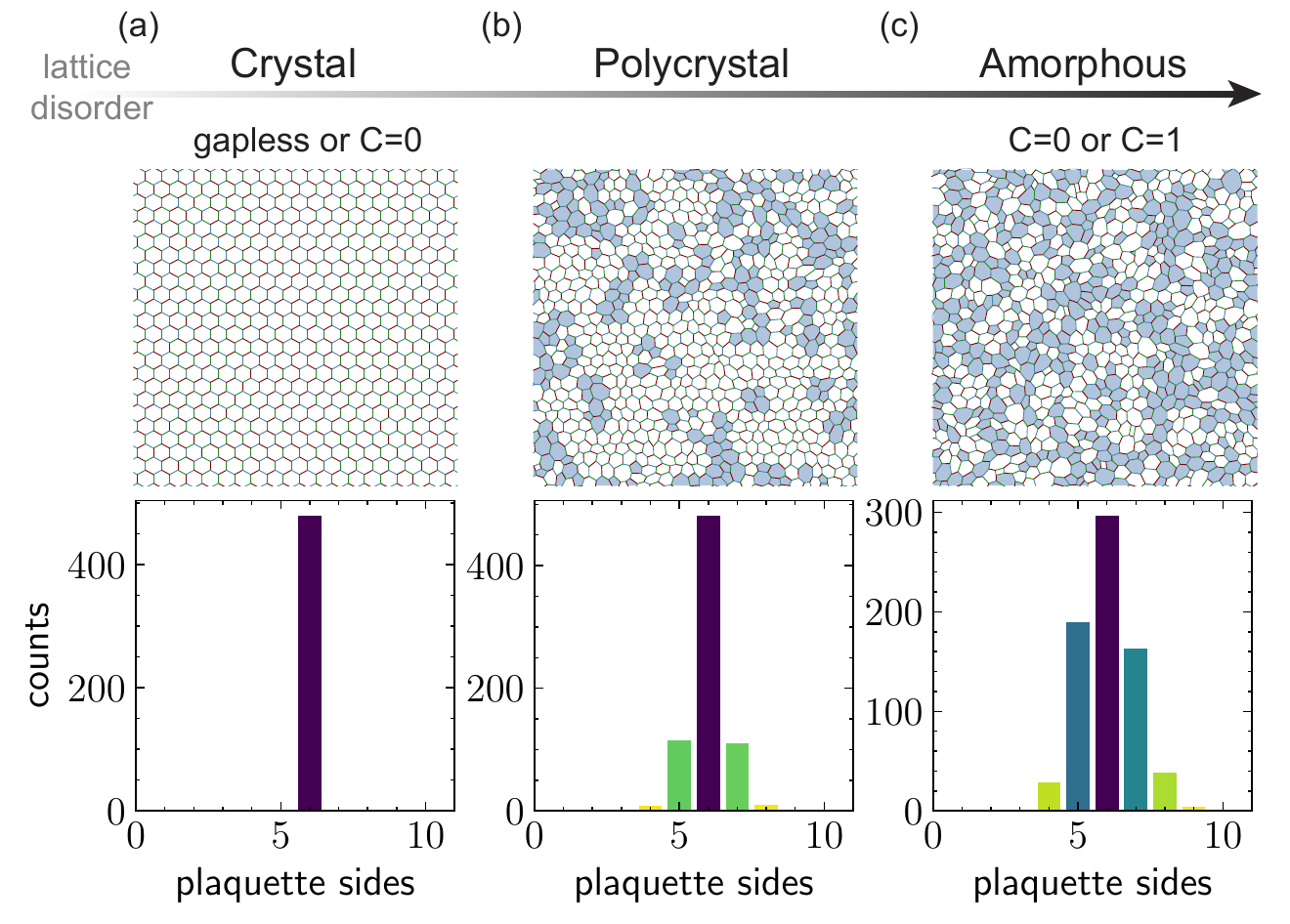}
    \caption{\label{fig:fig1} As the density of odd plaquettes, $n_{odd}=N_{odd}/N_{tot}$, is increased from (a) $n_{odd}=0$ to (b) $n_{odd}=0.36$ to (c) $n_{odd}=0.49$, the honeycomb Kitaev model undergoes a transition from a gapless, or gapped spin-liquid with Chern number $C=0$, into a gapped phase with Chern numbers $C=0$ or $C=\pm 1$ and broken time-reversal symmetry.
    }
\end{figure}

The paradigmatic Kitaev honeycomb model~\cite{Kitaev:2005ik} is a good starting point to find amorphous spin-liquid candidates. First, Kitaev material Li$_2$IrO$_3$ has already been grown amorphous~\cite{Lee2019}, and more materials~\cite{jackeliPRL2009, Winter_2017,Takagi2019,Hermanns2018, TREBST20221} could soon follow, as nearly all crystals can be grown amorphous~\cite{Zallen}. Second, the three-colored Kitaev spin-$1/2$ model can be defined on any (crystalline or amorphous) three-coordinated lattice and remains exactly solvable through a mapping to a model of free Majorana fermions coupled to a static $\mathbb{Z}_2$ gauge field. Third, amorphous materials may facilitate the observation of a chiral spin liquid by removing the need for a magnetic field. Indeed, as already noted by Kitaev~\cite{Kitaev:2005ik}, in the presence of odd plaquettes (plaquettes with an odd number of edges), the ground state spontaneously breaks time-reversal symmetry. This occurs in the decorated honeycomb lattice~\cite{Yao:2007jd, dusuel-prb2008}, the pentaheptite tiling~\cite{Peri2020}, and in the amorphous graphene lattice~\cite{Cassella2022}, which all host a finite density of odd plaquettes. 

These observations suggest that there exists an unexplored and advantageous phenomenology to engineer a chiral spin liquid in amorphous and polycrystalline materials (see Fig.~\ref{fig:fig1}). To take advantage of it, we need to determine the minimal amount of structural disorder, {or density of odd plaquettes}, needed to  obtain a chiral spin liquid, and realistic methods to engineer a density of defects experimentally.

In this work, we propose experimental pathways to realize a chiral spin liquid in the laboratory. We first determine the amount of amorphicity required to transform a Kitaev spin liquid into a gapped chiral spin liquid. We find that the gap linearly increases with the density of odd plaquettes, $n_{odd}$, saturating at a value of $0.1$ times the Kitaev exchange coupling $J$ at $n_{odd}\sim 40\%$. 
Since threefold coordinated amorphous materials, such as amorphous graphene~\cite{Toh:2020dy}, typically experimentally exhibit a density of odd plaquettes of $n_{odd}\sim 30\%$, a sizable gap seems well within experimental reach. Moreover, we find that Kitaev materials grown as polycrystalline samples with a sufficiently homogeneous spread of defects (as in Fig.~\ref{fig:fig1}(b)), or samples patterned with a focused-ion beam could realize a chiral spin liquid. We also determine the stability of the chiral spin liquid to Heisenberg nearest-neighbor interactions, which are expected to be present in Kitaev materials. We find that amorphicity does not significantly change the stability of the Kitaev spin liquid, i.e. the amorphous chiral spin liquid is almost as stable as its crystalline gapless spin-liquid counterpart.

\begin{figure}
    \centering
    \includegraphics[width=\columnwidth]{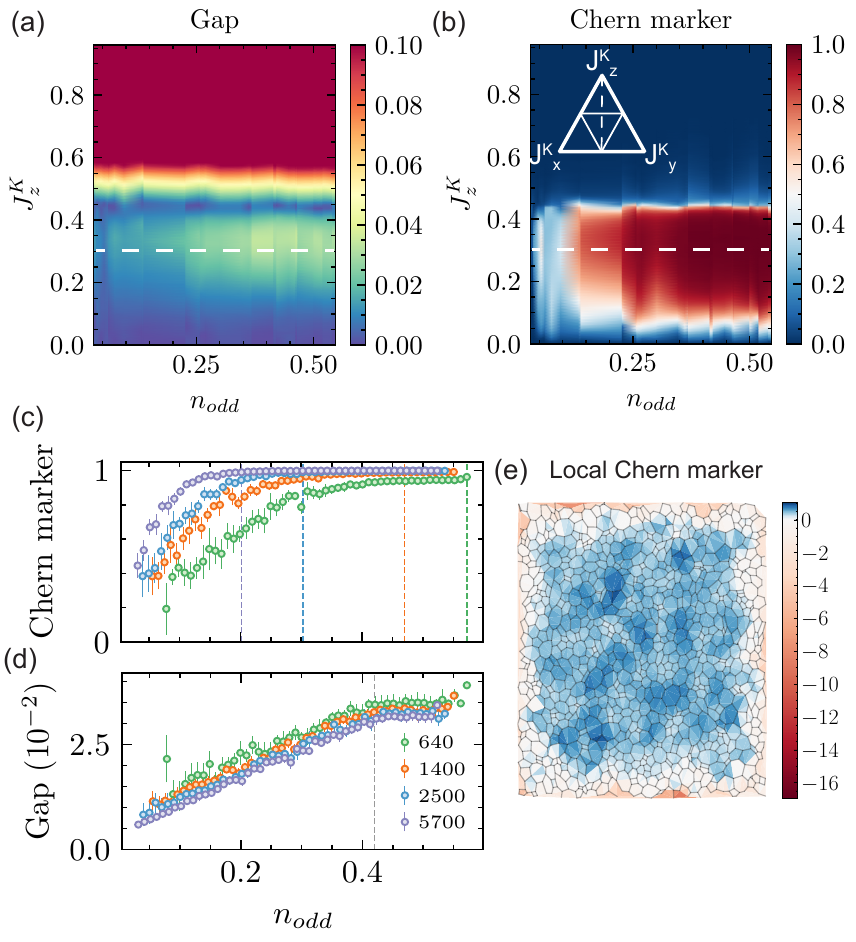}
    \caption{\label{fig:fig_pd}Phase diagram showing the magnitude of the (a) gap in units of $J^{K}_z$  and (b) Chern marker center value $C$ as a function of the coupling $J^{K}_z$ and the fraction of odd plaquettes $n_{odd}=N_\mathrm{odd}/N_{\mathrm{tot}}$. The parameters satisfy $J^{K}_x=J^{K}_y=(1-J^{K}_z)/2$, and define a vertical cut of the phase diagram of the crystalline Kitaev model (vertical dashed line in triangle inset of (b)). Taken together, (a) and (b) indicate a chiral gapped spin-liquid phase with $C=1$ on the lower right of the phase diagram. Both plots are calculated for a Voronoi seed of size $L_x=L_y=30$ (see main text), resulting in $\sim 1400$ sites and $\sim 900$ plaquettes. (c) and (d) show a $J^{K}_z=0.3$ cut in the phase diagrams of (b) and (a), respectively, as a function of system size (horizontal dashed line in (a) and (b)). The vertical axis values are computed from 20 disorder realizations for each $w$, by binning the resulting $n_{odd}$ axis in $60$ bins and taking the average within each bin. The bin center gives the horizontal value, $n_{odd}$. The standard deviations for each bin are shown as error bars. The vertical dashed lines in (c) indicate the $n_{odd}$ after which $C$ is within $1\%$ of the quantized value. The vertical dashed line in (d) indicates the approximate $n_{odd}\approx 0.4$ after which the gap saturates. (e) Local Chern marker~\cite{Bianco2011} deep in the chiral spin-liquid phase, with a bulk quantized value of $C=1$.}
\end{figure}

We start with the Kitaev model~\cite{Kitaev:2005ik} defined on a lattice of coordination three,
\begin{equation}
\label{eq:model}
    H_k =  \sum_{i,j,\alpha} J^K_{\alpha} \sigma^{\alpha}_i \sigma^{\alpha}_j,
\end{equation}
where $\sigma^{\alpha=x,y,z}_j$ is the spin-$1/2$ Pauli operator acting on site $j$, and the nearest-neighbor bonds $\alpha$ satisfy the three coloring of the lattice. The flux operator $W_p=\Pi \sigma_{i}\sigma_{k}$ is defined on each plaquette as the product over all bonds at its boundary; all $W_p$ commute with $H_k$ and with themselves. The ground state is found in the flux sector $\phi_p$ such that $\phi_p=-(\pm i)^{n_\mathrm{sides}}$, as recently shown in Ref.~\onlinecite{Cassella2022}. 
On the honeycomb lattice, the phase diagram is conventionally represented by the triangle where $\sum_{\alpha}J^{K}_{\alpha} = 1$, with $J^{K}_\alpha>0$ (see inset of Fig. \ref{fig:fig_pd}(b)). At the center of the triangle $J^{K}_{x}=J^{K}_{y}=J^{K}_{z}$, and the model realizes a gapless spin liquid. In contrast, when one of the $J_{\alpha}$ dominates, the ground state is gapped and preserves time-reversal symmetry, as the Chern number of the underlying Majorana model is $C=0$.

We now solve this model in lattices with different densities of odd plaquettes $n_{odd}=N_{\mathrm{odd}}/N_{\mathrm{tot}}$ and $J^{K}_\alpha$. Each lattice is generated by voronization of a point set~\cite{Marsal2020}. The voronization procedure finds the area closest to a given point of the point set. The edges and vertices of each area define the edges and vertices of a threefold coordinated lattices. For a perfect triangular point set of size $L_x\times L_y$, the voronization procedure produces a perfect hexagonal lattice. By displacing the points of a triangular point set, with a probability drawn from a normal distribution with standard deviation $w$, we can generate lattices with different values of $n_{odd}$, controlled by $w$, as seen in Fig.~\ref{fig:fig1}. For a given lattice the tree-coloring is then implemented using the algorithm of Refs.~\cite{Cassella2022,tom_2022}.

The density $n_{odd}$ and the anisotropy of the coupling terms $J_{\alpha}$ determines the phase and gap size of the ground state of Eq.~\eqref{eq:model}. We show in Fig. \ref{fig:fig_pd} (a) and (b) the gap and Chern number diagram of the model for different values of $J^{K}_x=J^{K}_{y}=(1-J^{K}_z)/2$ and $n_{odd}$. The gap is computed by solving numerically \eqref{eq:model} in the Majorana representation with periodic boundary conditions. Because the system lacks translational invariance, we compute the real-space local Chern marker~\cite{Bianco2011,Marsal2020,Ornellas2022} in the bulk of the system,  $C$, of the quadratic Majorana Hamiltonian corresponding to each disorder realization, using the method of Refs.~\cite{Ornellas2022,tom_2022}.

We focus on a line with $J^K_x=J^K_y$ and $J^K_x+J^K_y+J^K_z=1$, which splits the triangle phase diagram in two (see inset of Fig.~\ref{fig:fig_pd}b). For large enough values of $J^{K}_z$ ($J^{K}_z \gtrsim 0.4$), we find that the ground state is gapped with $C=0$ for any value of $n_{odd}$. Below this threshold, two different phases are possible depending on the value of $n_{odd}$. When $n_{odd}=0$, the ground state of the Kitaev model is a gapless spin liquid. As $n_{odd}$ increases, a chiral spin liquid phase with $C=1$ for the underlying Majorana fermions appears. This phase is enabled by a finite density of odd plaquettes and it is unique to adding structural disorder. {Our calculation of the local spin scalar chirality~\cite{SuppMat} confirms that time-reversal invariance is broken locally around the odd plaquettes.}

In Fig.~\ref{fig:fig_pd} (c) and (d) we fix $J^{K}_z=0.3$ and plot the disorder-averaged bulk Chern marker $C$ and the gap versus $n_{odd}$, respectively, for different system sizes. For all system sizes, the gap in (d) shows a linear increase with $n_{odd}$, saturating around to a gap $\simeq 0.03$ above $n_{odd} \simeq 0.4$. The onset of the Chern marker quantization in (c) shifts to smaller and smaller $n_{odd}$ as the system size is increased. This is apparent by the shift of the vertical dashed lines, which indicate the value of $n_{odd}$ above which the Chern marker is quantized within $1\%$. 
For lower $n_{odd}$, larger system sizes are required to reach the thermodynamic limit, and thus to reach exact quantization of the Chern marker, due to the smaller gap (larger correlation length).

Our results show that a very small critical density of odd plaquettes $n^c_{odd} \lesssim 0.05$ is sufficient to open a gap above the spin-liquid ground state in the thermodynamic limit. Numerically, it is challenging to determine whether $n^c_{odd}$ is zero or very small, {and our results do not rule out a zero $n^c_{odd}$}. The accuracy of $n^c_{odd}$ is limited by the difficulty to access very small ($\lesssim 0.05$) values of $n_{odd}$, since  our Voronoi procedure only leads to small variations of $n_{odd}$ at small $w$.

Overall, our findings presented in Fig.~\ref{fig:fig_pd} suggest that a relatively low density of odd plaquettes is needed to turn a honeycomb lattice Kitaev spin-liquid into a chiral Kitaev spin-liquid. Such situation may be realized naturally in amorphous versions of Kitaev honeycomb materials. For example, amorphous graphene samples~\cite{Toh:2020dy}, whose crystalline phase is also hexagonal, shows regions of hexagons coexisting with regions with odd plaquettes, mostly heptagons and pentagons as in Fig.~\ref{fig:fig1} (b). It is thus likely that a similar level of structural disorder can occur in amorphous Kitaev materials.

\begin{figure}
    \centering
    \includegraphics[width=\columnwidth]{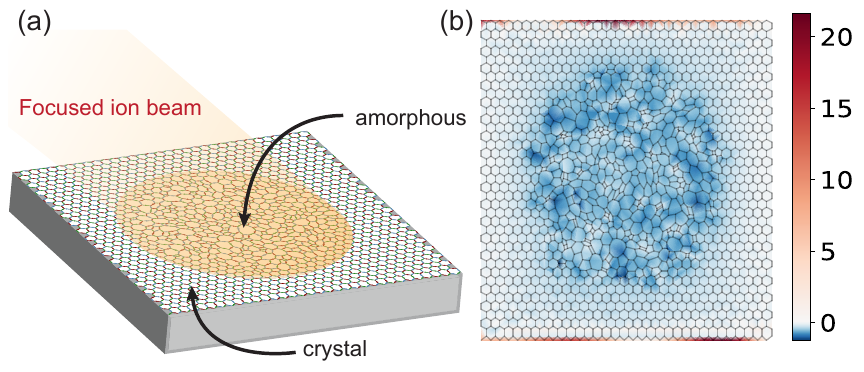}
    \caption{\label{fig:fig_fib} Engineering a chiral spin-liquid: (a) A focused ion beam can turn an crystal region into an amorphous solid. If the underlying sample is a Kitaev spin-liquid candidate, this method can turn the irradiated part of the system into a chiral spin liquid, signalled by a finite local Chern marker (b). The couplings in (b) are chosen to be $J^{K}_x=J^{K}_y=J^{K}_z=1$. }
    \end{figure}

A more controlled possibility is to create an amorphous region that is above the $1/4$ density threshold in an otherwise trivial Kitaev material. Specifically, a focused ion beam can be used to amorphisize a region of a given sample Fig.~\ref{fig:fig_fib}(a). Computing the local Chern marker, Fig.~\ref{fig:fig_fib}(b), we observe that indeed the Chern number is quantized to $C=-1$ in the central region, indicating a defect induced chiral spin-liquid phase.

In candidate Kitaev materials, the Kitaev interaction \eqref{eq:model} comes in addition to other spin interaction terms, predominantly the nearest-neighbor Heisenberg interaction~\cite{jackeliPRL2009} (see also Refs.~\onlinecite{Winter_2017,Takagi2019, Hermanns2018, TREBST20221} and references within). This gives rise to the Kitaev-Heisenberg Hamiltonian 
\begin{equation}
\label{eq:heis}
    H_{kh} = J^K\sum_{i,j,\alpha} \sigma^{\alpha}_i\sigma^{\alpha}_j  +J^H\sum_{\left\langle ij \right\rangle} \boldsymbol{\sigma}_i \cdot \boldsymbol{\sigma}_j,
\end{equation}
where we have set $J^K_x = J^K_y = J^K_z = J^K$.
Unlike the pure Kitaev model~\eqref{eq:model}, Eq.~\eqref{eq:heis} does not map to a problem of noninteracting Majorana fermions. The Kitaev-Heisenberg phase diagram has been established on the honeycomb lattice~\cite{chaloupkaPRL2010, chaloupkaPRL2013}, showing that the Kitaev spin liquid is surrounded by various phases with long-range order. We can hope that the amorphous lattice will frustrate these ordered phases, thus increasing the stability of the spin liquid. In the absence of Heisenberg interactions ($J^H=0$), the amorphous ground state is a chiral spin liquid~\cite{Cassella2022} regardless of the sign of $J^K$ due to the particle-hole symmetry of the Majorana Hamiltonian. We now investigate its phase boundaries, and we leave the investigation of the full Kitaev-Heisenberg phase diagram to future studies. 

\begin{figure}
    \centering
    \includegraphics[width=\columnwidth]{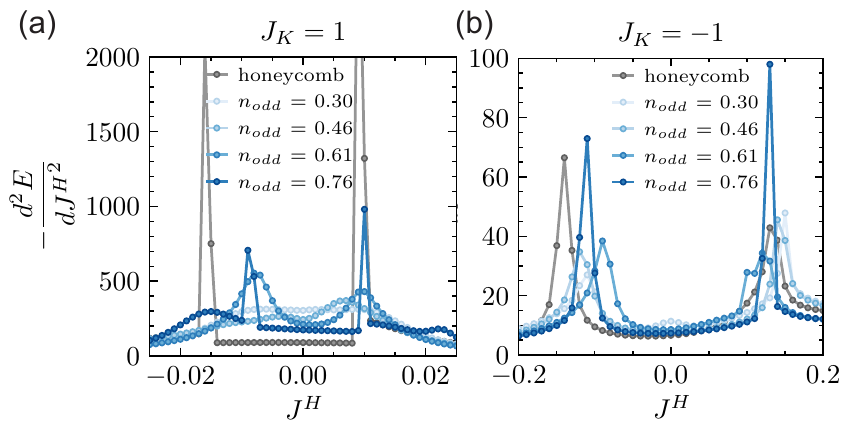}
    \caption{\label{fig:heis} {Phase boundaries of the amorphous Kitaev spin liquid upon adding a Heisenberg term $J^H$, for (a) $J_{K}=1$ and (b) $J_{K}=-1$}, as probed by the second derivative of the ground state energy $d^2E/dJ_H^2$ in a system of $26$ spins (13 plaquettes). Four random amorphous realizations are shown, with increasing number of odd plaquettes (4, 6, 8, and 10, in shades of blue), in addition to the honeycomb system for the same system size (gray). The phase transitions are signaled by a divergence of $d^2E/dJ_H^2$.}
    \end{figure}

To determine the stability of the amorphous Kitaev spin liquid, we use exact diagonalization of finite-size clusters with up to $26$ spins and periodic boundary conditions. As is typical in the numerical studies of the Kitaev-Heisenberg model pioneered in Ref.~\onlinecite{chaloupkaPRL2010}, we calculate the second derivative of the ground state energy, whose singularities indicate the position of the phase transitions. Our results, summarized in Fig.~\ref{fig:heis}, indicate that the stability of the amorphous chiral spin liquid depends on the signs of $J^K, J^H$, and is either the same or somewhat smaller than the stability of the honeycomb Kitaev spin liquid.  In the case of ferromagnetic Kitaev interactions ($J^K=-1$), we find that the spin liquid is stable up to $J^H \simeq \pm 0.12$. Similar to the honeycomb geometry, the $J^K=1$ spin liquid appears less stable, up to $J^H \simeq \pm 0.01$. Yet, our $J^K=1$ results are harder to analyze due to larger finite-size effects. Indeed, different amorphous realizations show qualitatively different behaviors, and the phase boundaries of the honeycomb lattice still vary by a factor of two between the two largest system sizes we have looked at ($24$ and $26$ spins). Additional exact diagonalization results, including results for $24$ spins, are presented the Supplemental Material \cite{SuppMat}.

We now focus on the ferromagnetic Kitaev regime ($J^K=-1$), where our numerical results are more reliable.
Overall, the spin liquid ground state appears similarly robust to Heisenberg interactions in geometries as different as the honeycomb and amorphous lattices. This is especially surprising since one ground state (honeycomb) is gapless, while the other (amorphous) is gapped. Explaining the quantitative value of $J_H/J_K$ at the phase transition is a challenging task beyond the scope of this work. Yet, we note that for a large enough $-J_H$, a ferromagnetic ground state is expected irrespective of the lattice geometry, which may explain some similarities. The situation is more subtle for $J_H>0$, where the {ground state has} stripy long-range order on the honeycomb lattice, and cannot be easily generalized to the amorphous lattice.

To gain insight into the amorphous ground state at $J^H > 0, J^K < 0$ we focus on the Hamiltonian $J^H = -J^K/2$, which admits an exact stripy ground state on the honeycomb lattice. The Hamiltonian reads
\begin{equation}
\label{eq:kit-heis}
    H_{kh}\left(J^H = -J^K/2 \right) = J^H \sum_{i,j} \left(-\sigma^{\alpha}_i\sigma^{\alpha}_j + \sigma^{\beta}_i\sigma^{\beta}_j + \sigma^{\gamma}_i\sigma^{\gamma}_j\right).
\end{equation}
On the honeycomb lattice, it is possible to apply a site-dependent spin rotation $\mathbf{\sigma} \rightarrow \mathbf{\tilde{\sigma}}$ such that the interaction becomes ferromagnetic on all bonds ($H_{kh} = -J_H\sum_{\left\langle ij \right\rangle}\boldsymbol{\tilde{\sigma}}_i \cdot \boldsymbol{\tilde{\sigma}}_j$)~\cite{Khaliullin_2005}, leading to an exact ferromagnetic  ground state in the rotated basis (stripy in the original spin basis). This rotation relies on the particular sequence of bonds in the Kitaev honeycomb model, and cannot be consistently performed on a generic amorphous lattice. We confirmed this intuition by checking the level spacing statistics of Eq.~\eqref{eq:kit-heis}'s energy spectrum; we found that it follows the Gaussian orthogonal ensemble distribution, as expected for a time-reversal symmetric Hamiltonian with no hidden conserved quantity\footnote{On the honeycomb lattice, the Hamiltonian Eq.~\eqref{eq:kit-heis} maps to a Heisenberg Hamiltonian, which has $SU(2)$ symmetry, and the GOE level spacing statistics is only recovered within one symmetry sector.}. Let us understand the impossibility to transform Eq.~\eqref{eq:kit-heis} into a Heisenberg Hamiltonian in the following. We start by picking a random initial spin on the lattice. Applying the appropriate spin rotation (flipping the sign of two out of three spin components) onto neighboring spins, we obtain three ferromagnetic bonds. The ferromagnetic tree can grow as long as one does not form a cycle, where the consistency cannot be guaranteed\footnote{Consider, e.g., a quadrilateral with the bond sequence x, y, x, z. We apply the identity onto the first spin, then $\tilde{\sigma^x} = \sigma^x, \tilde{\sigma^{y,z}} = -\sigma^{y,z}$ onto the second spin to obtain a ferromagnetic interaction on the first x bond. Applying the appropriate spin rotation sequentially onto each spin of the loop, we see that we can make all bonds ferromagnetic except the z bond. Which bond is not ferromagnetic depends on the choice of the initial transformation, but the number of frustrated bonds does not.}. Neglecting the number of cycle-closing bonds that are ferromagnetic by chance, the number of ferromagnetic bonds is $N_s -1$ (for a total of $3N_s/2$ bonds). This result comes from a theorem of graph theory~\cite{wikipedia:spanning_tree}, which states that connected graphs with $N_s$ vertices admit spanning trees (graphs with the same vertices as the original lattice, but no cycle) with $N_s-1$ edges. In the thermodynamic limit, the fraction of nonferromagnetic bonds is thus $1/3$ at most.  One possible variational state for the amorphous ground state at $J^H = -J^K/2$ may be a superposition of the ferromagnetic (in the rotated basis) states obtained for all possible trees of spin rotation (the number of such trees is expected to be exponential in the number of spins~\cite{combinatorics_graph}).

\paragraph*{Discussion}

Our work indicates that a chiral spin liquid spontaneously emerges in the Kitaev model for a small density of odd plaquettes. The resulting Majorana gap increases steadily with the density of odd plaquettes and reaches $75\%$ of the saturating value ($0.1J$) at defect densities found in amorphous solids ($n_{odd}\sim 30\%$).  Our results indicate that a chiral spin liquid, detectable for instance in thermal Hall conductance measurements~\cite{Kasahara2018,Yokoi2021,Yamashita2020,Bruin2022}, could be realized in amorphous and polycrystalline samples with a sufficiently homogeneous distribution of defects and three-fold coordination. The latter is physically plausible, as the local site environments of amorphous, polycrystalline and crystalline solids are dictated by the same local chemical rules~\cite{Zallen,Weaire:1971in,Marsal2020,Toh:2020dy}. These observations may have physical relevance for Kitaev candidate materials that can be grown amorphous, such as Li$_2$IrO$_3$~\cite{Lee2019}, and adds another element to the rich phenomenology of the interplay of disorder and spin liquids~\cite{Savary2017,Petrova2014,Brennan2016,Khan2017,Nasu2020}. Contrary to the naive expectation that amorphous lattices may frustrate long-range order, we found that amorphicity does not significantly change the stability of the Kitaev spin liquid with respect to Heisenberg interactions.

Additionally, we also proposed to use a focused ion beam  to add structural disorder the lattice. By locally creating an amorphous region, this method can trigger the formation of a chiral spin liquid phase, embedded in an otherwise crystalline and topologically trivial lattice.

We have also checked that the amorphous version of the decorated honeycomb has a chiral spin liquid ground state. However, since the ordered version also breaks time reversal due the presence of the decorating triangles~\cite{Yao:2007jd}, it is not surprising that the $C=1$ chiral spin-liquid survives amorphization. Once translational invariance is lost, the decorated amorphous lattices are a subclass of threefold coordinated amorphous lattices with a large density of triangles.

As future avenues, investigating the topological transition as a function of domain size, is worthy of further study.  
Specifically, the critical $n_{{odd}}$ needed to form an amorphous Chern insulator from a random collection of points\cite{Agarwala:2017jv,Sahlberg2020}, motivates the study of the percolation transition~\cite{Becker2009} along the lines of Ref.~\cite{Sahlberg2020}. 
Additionally, the effect of bond disorder should be considered when modelling realistic materials, as different bond lengths will lead to different spin interaction energy $J^{K}, J^H$. Previous studies in amorphous systems~\cite{Marsal2022} suggest that gapped topological phases in amorphous matter survive so long as the typical disorder strength is not sufficiently strong to close the mobility gap, a criterion similar to disordered, gapped crystalline phases. 
We leave a detailed study of these questions for future work.

\paragraph*{Acknowledgements} We thank G. Cassella, P. D'Ornellas, T. Hodson, J. Karel, W. M. H. Natori and J. Knolle for stimulating discussions. A.G.G. acknowledges financial support from the European Union Horizon 2020 research and innovation program under grant agreement No. 829044 (SCHINES) and the European Research Council (ERC) Consolidator grant under grant agreement No. 101042707 (TOPOMORPH).

\bibliography{a-kitaev.bib}

\clearpage
\newpage

\setcounter{secnumdepth}{5}
\renewcommand{\theparagraph}{\bf \thesubsubsection.\arabic{paragraph}}

\renewcommand{\thefigure}{S\arabic{figure}}
\setcounter{figure}{0} 

\appendix

\section{Additional exact diagonalization results for the Kitaev-Heisenberg model on amorphous lattices}

\begin{figure}[ht!]
    \centering
    \includegraphics[width=\columnwidth]{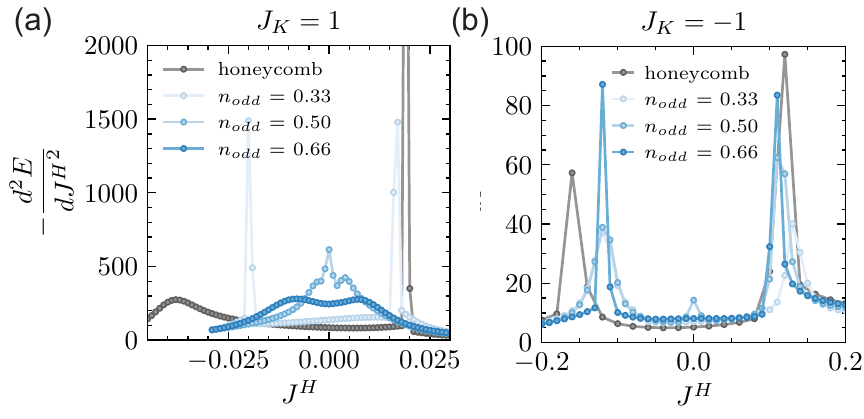}
    \caption{\label{fig:24} Phase boundaries of the Kitaev spin liquid in random amorphous lattices of $24$ spins (12 plaquettes), upon adding a Heisenberg term $J_H$ for positive (a) and negative (b) Kitaev interaction for different odd plaquette density (4, 6, and 8 plaquettes, in shades of blue). The phase transitions are signaled by a divergence of $d^2E/{dJ^H}^2$, the second derivative of the ground state energy. The phase boundaries on the honeycomb lattice with the same number of spins is also given for reference~\cite{chaloupkaPRL2013} (gray).}
    \end{figure}

We first specify the lattice geometry used for the (crystalline) honeycomb lattice with $26$ spins in Fig.~4 of the main text. To obtain $13$ unit cells on a torus with two equal-length cycles, we have used tilted periodic boundary conditions defined by $a_1 = e_1 + 3 e_2$ and $a_2 = -3 e_1 + 4 e_2$, where $e_1,e_2$ are the unit vectors of the triangular Bravais lattice.

In Fig.~\ref{fig:24}, we show the boundaries of the amorphous chiral spin liquid phase upon adding a Heisenberg interaction, for a system of $24$ spins, a smaller system than the $26$ spin system discussed in the main text. For $J^K < 0$, the phase boundaries of the amorphous systems are similar for both $24$ and $26$ spins. Their precise positions depend on the specific amorphous lattice, with similar variations $\Delta_{J^H}\simeq 0.05$ for both system sizes due to the amorphous realization. For $J^K>0$, these amorphous variations, as well as the small value of $|J^H|$ at the transition make it hard to distinguish the phase boundary. Moreover, the numerical results at $J^K>0$ are affected by much larger finite-size effects, already visible in the honeycomb geometry: the overall span of the Kitaev spin liquid is twice smaller for $26$ spins as for $24$ spins.

\begin{figure}[ht!]
    \centering
    \includegraphics[width=\columnwidth]{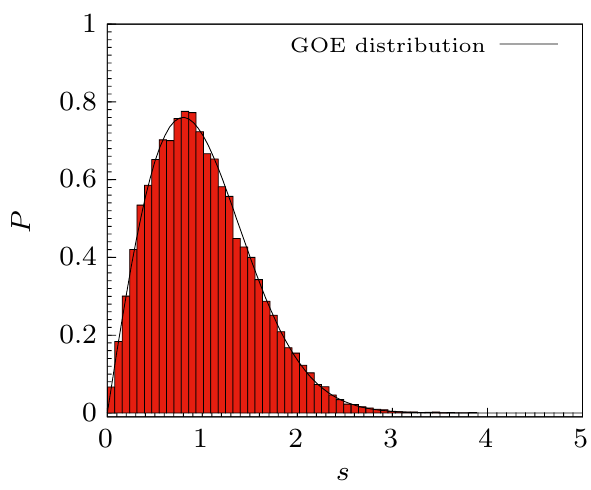}
    \caption{\label{fig:level spacing} Level spacing statistics of the amorphous Kitaev-Heisenberg energy spectrum with $J_H=0.5$, $J_K = -1$, shown here for a random amorphous system of $16$ spins. It falls on a gaussian orthogonal ensemble (GOE) distribution, as expected for a non-integrable system with time-reversal symmetry.}
    \end{figure}

In Fig.~\ref{fig:level spacing}, we show the level spacing statistics of the energy spectrum of the Kitaev-Heisenberg Hamiltonian Eq.~(2) of the main text. We call $N_\uparrow^p$ the parity of the total number of up spins, and $S_z^p$ the parity of the spin-flip operation (applying $\sigma_i^z \rightarrow -\sigma^z_i$ on all sites). $N_\uparrow^p$ and $S_z^p$ are both conserved quantities of Eq.~(2)of the main text for any value of $J^H$ and $J^K$, such that we can diagonalize the Hamiltonian in a reduced Hilbert space with fixed $N_\uparrow^p$ and $S_z^p$. The level spacings $s$ are obtained for the four spectra corresponding to four different values of $(N_\uparrow^p,S_z^p)$. Their distribution is then averaged over the four $(N_\uparrow^p,S_z^p)$ symmetry sectors to obtain Fig.~\ref{fig:level spacing}. The perfect agreement with the GOE distribution shows that there are no additional conserved quantities in this amorphous system.

{\section{Scalar spin chirality}
The expectation value of the three-spin scalar scalar chirality $\hat{\chi}_{ijk} = \mathbf{S}_i\cdot \left(\mathbf{S}_j \times \mathbf{S}_k \right)$ may be calculated in the ground state ($\ket{\psi_0}$) of spin models as a local and quantitative estimate of time-reversal symmetry breaking, especially in chiral spin liquids. In the Kitaev model, the chirality can be expressed in the Majorana language as~\cite{Kitaev:2005ik}
\begin{equation}
\langle \hat{\chi}_{ijk} \rangle 
=-i\bra{\psi_0}u_{ij}u_{kj}c_ic_k\ket{\psi_0},
\end{equation}
where $u_{ij}$ is the conserved bond operator and $c_i$ is a Majorana operator at site $i$.

Using the above expression, we have calculated $\langle \hat{\chi}_{ijk} \rangle$ in the ground state of the Kitaev model with $J^{K}_x=J^{K}_{y}=J^{K}_z$ on three different lattices: a) honeycomb lattice, b) honeycomb with a small density of odd plaquettes, and c) amorphous lattice. Our results, shown in Fig~\ref{fig:scalar chirality}, confirm that $\langle \hat{\chi}_{ijk}\rangle = 0$ everywhere on the honeycomb lattice (where time-reversal invariance is preserved), and becomes locally finite around the odd plaquettes of more general lattices, where time-reversal invariance is broken spontaneously.

\begin{figure*}
    \centering
    \includegraphics[width=\textwidth]{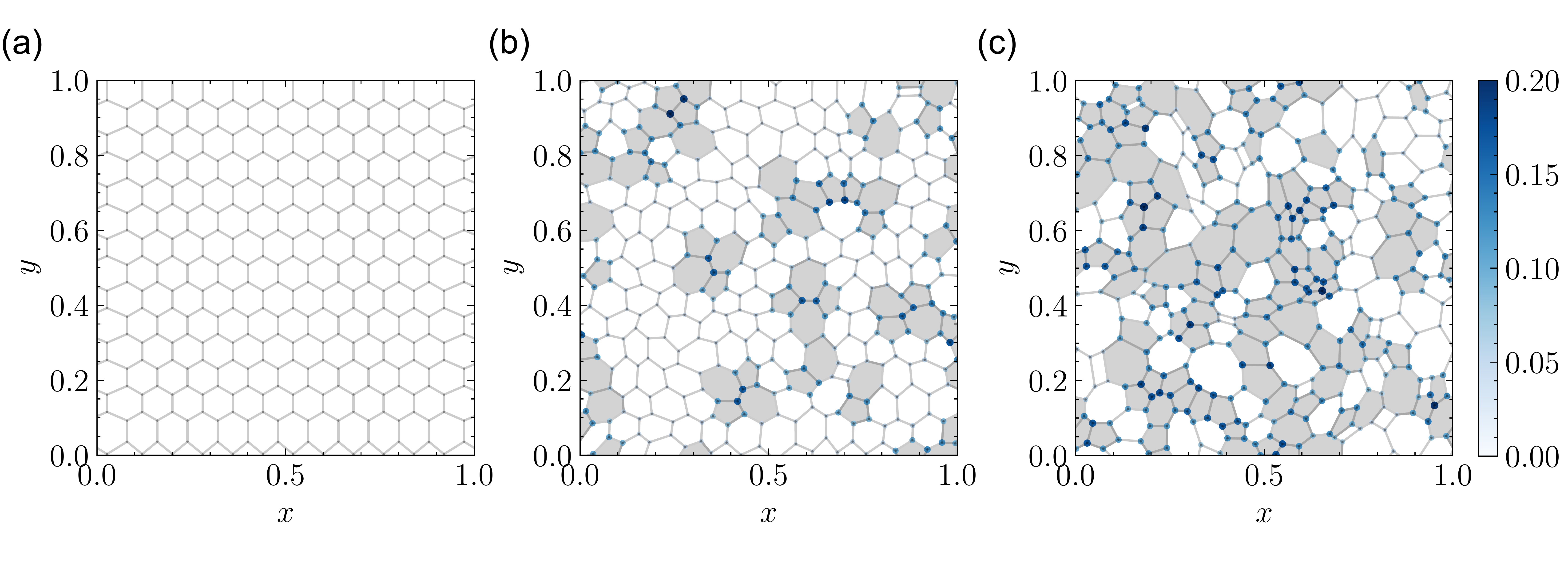}
    \caption{\label{fig:scalar chirality} Local expectation value of the three-spin scalar chirality in the ground state of the Kitaev model Eq.~(1). For ease of representation, we show the sum of the three contributions on each site: $\langle\hat{\chi_l}\rangle = \langle\hat{\chi}_{ilj}\rangle + \langle\hat{\chi}_{jlk}\rangle + \langle\hat{\chi}_{kli}\rangle$, where $i, j, k$ are $l$'s  three nearest neighbors. The size and color of each circle is proportional to the magnitude of $\langle\hat{\chi_l}\rangle$.}
    \end{figure*}
    }

\end{document}